\documentstyle[12pt]{article}

\global\arraycolsep=1pt
\newcommand{\beq}{\begin{equation}}
\newcommand{\eeq}{\end{equation}}
\newcommand{\beqa}{\begin{eqnarray}}
\newcommand{\eeqa}{\end{eqnarray}}
\newcommand{\CR}{\nonumber \\}

\newcommand{\del}{\partial}

\renewcommand{\theequation}{\thesection.\arabic{equation}}
\renewcommand{\thefootnote}{\fnsymbol{footnote}}

\begin{document}

\begin{titlepage}
\begin{flushright}
{August, 2001} \\
{\tt hep-th/0108226}
\end{flushright}
\vspace{0.5cm}
\begin{center}
{\Large \bf
On $Spin(7)$ holonomy metric based on $SU(3)/U(1)$
}%
\lineskip .75em
\vskip1.5cm
{\large Hiroaki Kanno\footnote{e-mail: kanno@math.nagoya-u.ac.jp}}
\vskip 1.5em
{\large\it Graduate School of Mathematics \\
Nagoya University, Nagoya, 464-8602, Japan}
\vskip1cm
{\large Yukinori Yasui\footnote{e-mail: yasui@sci.osaka-cu.ac.jp}}
\vskip 1.5em
{\large\it Department of Physics, Osaka City University \\
Sumiyoshi-ku, Osaka, 558-8585, Japan}
\end{center}
\vskip1.0cm

\begin{abstract}

We investigate the $Spin(7)$ holonomy metric of cohomogeneity one with
the principal orbit $SU(3)/U(1)$. A choice of $U(1)$
in the two dimensional Cartan subalgebra is left as free
and this allows manifest $\Sigma_3=W(SU(3))$ (= the Weyl group) symmetric
formulation. We find asymptotically locally conical (ALC)
metrics as octonionic gravitational instantons. These ALC metrics have
orbifold singularities in general, but a particular choice of the $U(1)$
subgroup gives a new regular metric of $Spin(7)$ holonomy.
Complex projective space ${\bf CP}(2)$ that is a supersymmetric four-cycle
appears as a singular orbit.
A perturbative analysis of the solution near the singular orbit shows
an evidence of a more general family of ALC solutions.
The global topology of the manifold depends on a choice of the $U(1)$ subgroup.
We also obtain an $L^2$-normalisable harmonic 4-form in the background
of the ALC metric.


\end{abstract}
\end{titlepage}

\renewcommand{\thefootnote}{\arabic{footnote}}
\setcounter{footnote}{0}

\section{Introduction}

In supersymmetric compactifications of
superstrings and $M$ theory the compact manifold must allow parallel spinors
and hence has a special holonomy.
Among manifolds of special holonomy, the holonomy
groups $G_2$ in seven dimensions and $Spin(7)$ in eight dimensions are
exceptional
ones. Recently compactification of $M$ theory on $G_2$ manifold has been
discussed
extensively in connection with $N=1$ supersymmetric gauge theory in four
dimensions \cite{Ach},\cite{AMV},\cite{Gom},\cite{EN},\cite{AW}.
Though it is less studied,
the geometry of $Spin(7)$ manifold is relevant to three
dimensional $N=1$ Yang-Mills theory.
Manifolds of exceptional holonomy with branes wrapping on a supersymmetric
cycle are also useful for establishing the gravity/gauge theory correspondence
that generalizes the AdS/CFT correspondence. In addition to few basic examples
of $G_2$ and $Spin(7)$ metrics on the total space of vector bundles
\cite{BS},\cite{GPP},
we now have an increasing list of explicit metrics
\cite{CGLP1},\cite{CGLP2},\cite{CGLP3},\cite{CGLP4},\cite{CGLLP},\cite{BGGG},\cite{KN}.
All these examples are metrics on non-compact manifolds
and of cohomogeneity one. There is a (rigid) supersymmetric cycle
and the non-compact manifold may be identified as the normal
bundle of the supersymmetric cycle.  In studying the dynamics of
supersymmetric compactifications of superstring
and $M$ theory, we are especially interested in the behavior when the
manifold develops singularities. Potentially there are various types of
singularities, but an important class of singularities
in supersymmetric dynamics is the isolated conical singularity
that arises when a supersymmetric cycle is shrinking.
In such cases the stringy geometry
is believed to be governed by a tubular neighborhood of the singularity or
the shrinking supersymetric cycle,
where the above explicit metrics on the normal bundle are available.
Furthermore, the geometry of such metrics often shows some universal feature
that is independent of the way singularities or supersymmetric cycles are
embedded in a manifold of special holonomy.
The geometry of $ADE$ singularities in $K3$ surface and
the conifold transition in Calabi-Yau threefold are typical examples
and we expect it is also the case with exceptional holonomy.

Let us review the basic geometry of manifolds of cohomogeneity one,
following \cite{DS}, \cite{DW}, \cite{EW}.
A Riemannian manifold $(M,g)$ is called cohomogeneity one,
if there is an isometric action on $M$ of a Lie group $G$ with generic orbit
of codimension one. The generic orbit $G/K$ is called principal orbit.
There is an open interval $I$ in real numbers with coordinate $t$, such that
$\widetilde M := I \times G/K$ is an open dense subset in $M$. The
compliment of
$\widetilde M$ consists of singular orbits of lower dimensions, where
we have a larger isotropy subgroup $H,~(K \subset H \subset G)$.
A tubular neighborhood of the singular orbit $Q=G/H$ is diffeomorphic
to an open disk bundle of the normal bundle of $Q$. Then the principal orbits
are the hypersurfaces which are the sphere bundles over $Q$. This
means $H/K$ is diffeomorphic to a sphere $S^k$. Thus, as the radius of the
sphere tends to zero, the principal orbits collapse to the singular orbit.
Furthermore, the existence of a smooth complete metric on the normal
bundle implies that the singular orbit must be a minimal submanifold.
We see the metric of cohomogeneity one is well suited
for describing the geometry of collapsing
supersymmetric cycles by identifying its normal bundle with
a manifold of cohomogeneity one.
To find out explicit metrics we begin with the fact that
on $\widetilde M$ the metric $g$ takes the following form;
\beq
g = dt^2 + g_t~,
\eeq
where the interval $I$ becomes a geodesic line. For each fixed ``time" $t$,
$g_t$ is a homogeneous metric of the principal orbit $G/K$. Hence
if we assume that the metric is of cohomogeneity one, the condition of
Ricci-flatness, or the Einstein equation in general, is reduced to
a system of non-linear ordinary differential equations with respect to
the transverse coordinate $t$ to the principal orbit.

In this paper we consider eight dimensional metrics of cohomogeneity one
with the principal orbit $SU(3)/U(1)$.  Part of our analysis is quite
parallel to
the case with the principal orbit $Sp(2)/Sp(1)$ which has been worked out in
\cite{CGLP3},
but there is a new feature that arises from the Weyl group symmetry
$\Sigma_3=W(SU(3))$.
We shall pay attention to this symmetry. In section two we derive
a first order system of
non linear differential equations from the octonionic self-duality
of the spin connection. If we choose vielbein (or metric ansatz) appropriately,
the octonionic self-duality of the spin connection implies
an existence of covariantly constant four form
which characterizes $Spin(7)$ holonomy. We also show that there is
a superpotential which implies the first order system.
In section three we present special solutions which give
asymptotically locally conical (ALC) \cite{CGLP3} metrics.
Our ansatz for special solution was motivated by the one in \cite{BGGG}.
Compared with ALC solutions in \cite{CGLP3} and \cite{BGGG},
our solution takes more general form to keep
the Weyl group $\Sigma_3$ symmetry manifest.
The singular orbit of $SU(3)/U(1)$ model
is the complex projective space ${\bf{CP}}(2) = SU(3)/U(2)$ which is self-dual
Einstein but not spin.
This is in a sharp difference from the $Sp(2)/Sp(1)$ case whose singular
orbit is the four dimensional sphere $S^4 = Sp(2)/Sp(1) \times Sp(1)$
which is self-dual Einstein and spin \cite{BS}.
Thus the issue of global topology is more subtle in our case.
We make a perturbative analysis around the singular orbit in section four
and find one more parameter in addition to the scale parameter in
the explicit ALC solutions in section three. This additional parameter
is an evidence for the existence of non-trivial deformation of our
ALC metrics and numerical simulations support it.
In section five we  discuss the global topology that depends on a choice
of the embedding of $U(1)$ subalgebra.
In general the fiber over the singular orbit ${\bf{CP}}(2)$ is the
Lens space $S^3/{\bf Z}_n$ which leads to orbifold singularities.
But there is a particular choice of $U(1)$ embedding which is free
from orbifold singularities. With this choice of $U(1)$ subalgebra,
our ALC solution gives a new $Spin(7)$ metric on a vector bundle
over  ${\bf{CP}}(2)$. Finally section six is devoted
to the construction of $L^2$-normalisable harmonic 4-forms in the background
of ALC metrics, which we can employ in constructing supersymmetric
M2-branes \cite{CGLP2},\cite{CGLP3},\cite{CGLP4}.

\section{Octonionic Instanton Equation}
\setcounter{equation}{0}

We consider an eight dimensional metric of cohomogeneity one with
the principal orbit $SU(3)/U(1)$. It is convenient to describe
homogeneous metric in terms of the Maurer-Cartan forms of $SU(3)$.
The Maurer-Cartan equation is presented in Appendix A.
We take a basis $T_A, T_B$ of the Cartan part and
$\sigma_{1,2}, \Sigma_{1,2}, \tau_{1,2}$ of non-Cartan part.
The Weyl group of $SU(3)$ is the permutation group $\Sigma_3$ and
our basis is chosen so that the Marter-Cartan eqaution
exhibits $\Sigma_3$ symmetry.
The isotropy representation of $SU(3)/U(1)$ is decomposed as
\beq
su(3)/u(1)={\bf p}_1 \oplus {\bf p}_2 \oplus {\bf p}_3 \oplus
{\bf p}_4, \label{dec}
\eeq
where ${\bf p}_i \; (i=1 \sim 4)$ are irreducible $U(1)$-modules
with $\mbox{dim}~{\bf p}_1=
\mbox{dim}~{\bf p}_2=\mbox{dim}~{\bf p}_3=2$
and $\mbox{dim}~{\bf p}_4=1$.
Our metric ansatz is diagonal with respect to (\ref{dec})
for all $t$~;
\beq
g = dt^2 + a(t)^2 (\sigma_1^2 + \sigma_2^2) + b(t)^2
        (\Sigma_1^2 + \Sigma_2^2) + c(t)^2 (\tau_1^2 + \tau_2^2)
+ f(t)^2 T_A^2~. \label{metric}
\eeq
We have taken a quotient by the $U(1)$ subgroup generated by $T_B$.
The vielbein of the above metric is
\beqa
& & e^0 = dt~, \quad e^1 = a(t) \sigma_1~,\quad e^2 = a(t) \sigma_2~,
\quad e^3 = b(t) \Sigma_1~, \CR
& & e^4 = b(t) \Sigma_2~,
\quad e^5 = c(t) \tau_1~, \quad e^6 = c(t) \tau_2~,\quad e^7 = f(t) T_A~.
\eeqa
The spin connection $\omega_{ab}$ is obtained from the condition
$De^a = de^a + \omega_{ab} \wedge e^b = 0$. We consider the octonionic
self-duality
of the spin connection
\beq
\omega_{ab} = \frac{1}{2} \Psi_{abcd} \omega_{cd}~, \label{octduality}
\eeq
where totally anti-symmetric tensor $\Psi_{abcd}$ is defined by
the structure constants of octonions $\psi_{abc}$ as follows;
\beqa
\Psi_{abc0} &=& \psi_{abc}~,  \quad (1 \leq a,b,c, \cdots \leq 7) \CR
\Psi_{abcd} &=& -\frac{1}{3!} \epsilon_{abcdefg} \psi_{efg}~.
\eeqa
It can be shown that (\ref{octduality}) implies the four form defined by
\beq
\Omega = \frac{1}{4!}
\Psi_{abcd} e^a \wedge e^b \wedge e^c \wedge
e^d~. \label {calib}
\eeq
is closed and the metric has $Spin(7)$ holonomy \cite{BFK}.
Explicitly the octonionic instanton equation in the present case is given by
the structure constants\footnote{An appropriate permutation of the indices and
the overall parity (sign) change are necessary to match our convention to
the standard one. This parity change is an analogue of the exchange of
self-duality and anti-self-duality in four dimensions and related to the
orientation of
the manifold.};
\beqa
\psi_{abc}= +1~, ~~~{\rm for}~~~(abc)=(721), (641), (135), (254),(263),
(374),(765)~.
\eeqa
We obtain the following first order differential equations;
\beqa
\frac{\dot a}{~a~} &=& \frac{b^2 + c^2 - a^2}{abc} - \alpha_A
\frac{f}{a^2}~, \CR
\frac{\dot b}{~b~} &=& \frac{c^2 + a^2 - b^2}{abc} - \beta_A
\frac{f}{b^2}~, \label{inst} \\
\frac{\dot c}{~c~} &=& \frac{a^2 + b^2 - c^2}{abc} - \gamma_A
\frac{f}{c^2}~, \CR
\frac{\dot f}{~f~} &=& \alpha_A \frac{f}{a^2}
+ \beta_A \frac{f}{b^2} + \gamma_A \frac{f}{c^2}~, \nonumber
\eeqa
where the parameters $\alpha_A, \beta_A, \gamma_A$ appearing in the Maurer-Cartan
equation of $T_A$ satisfy the ``traceless" condition $\alpha_A + \beta_A
+\gamma_A=0$.
These parameters have to be rational for the $U(1)$ subgroup generated by $T_B$
to be a closed subgroup (see also section 5).
We assume this condition required by topological consistency in the following.
Then there exists an integer $N$ so that $
(\alpha_A,\beta_A,\gamma_A) = (1/N)  (n_1, n_2, n_3)$ and
$\overrightarrow n:=(n_1, n_2, n_3)$ are integers with no common divisor.
Since $N$ is eliminated by the rescaling $f \longrightarrow N f$,
we may assume $(\alpha_A,\beta_A,\gamma_A) = (n_1, n_2, n_3)$
without any loss of generality. Our $Spin(7)$ gravitational instanton
equation is manifestly symmetric under the permutation group
$\Sigma_3=W(SU(3))$,
which can be regarded as the Weyl group of $SU(3)$.

We can also derive the octonionic instanton equation (\ref{inst}) from
the Lagrangian formulation. In the description of the extrinsic
geometry of hypersurface, the shape operator ${\cal L}$ of the
principal orbit $SU(3)/U(1) \subset \widetilde M$ satisfies
the equation \cite{EW}
\beq
\dot{g_t} = 2g_t \circ {\cal L}~.
\eeq
For the metric (\ref{metric}) it has a diagonal form,
\beq
{\cal L}(t) = \mbox{diag} \left(\frac{\dot{a}}{a}, \; \frac{\dot{a}}{a}, \;
\frac{\dot{b}}{b}, \; \frac{\dot{b}}{b}, \; \frac{\dot{c}}{c}, \;
\frac{\dot{c}}{c}, \; \frac{\dot{f}}{f} \right).
\eeq
The Ricci-flatness condition then becomes \cite{EW}
\beqa
\dot{{\cal L}} + (\mbox{tr}{\cal L}){\cal L} - Ric = 0~, \label{ric1} \\
\mbox{tr}\dot{{\cal L}} + \mbox{tr}({\cal L}^2) = 0~, \label{ric2}
\eeqa
where $Ric$ denotes the Ricci curvature of the metric $g_t$ on
$SU(3)/U(1)$. The equation (\ref{ric1}) expresses the Ricci-flatness
condition in directions tangent to the principal orbit, while
(\ref{ric2}) is obtained by considering the normal direction,
i.e., $t$-direction. The Ricci flatness of the mixed directions
is automatically satisfied.
This system of non-linear differential equations is described by
the Lagrangian $L=T-V$;
\beqa
T &=& ((\mbox{tr}{\cal L})^2 - \mbox{tr}({\cal L}^2))
\sqrt{\mbox{det}g_t}~, \CR
V &=& -R \sqrt{\mbox{det}g_t}~,
\eeqa
where $\mbox{det}g_t=a^4 b^4 c^4 f^2$ and
$R$ is the scalar curvature of $g_t$. After some calculation
we find
\beq
R=-2\left(\frac{a^2}{b^2 c^2}+\frac{b^2}{a^2 c^2}+\frac{c^2}{a^2 b^2}
-\frac{6}{a^2}-\frac{6}{b^2}-\frac{6}{c^2} \right)
-2f^2 \left(\frac{n_1^2}{a^4}+\frac{n_2^2}{b^4}
+\frac{n_3^2}{c^4} \right).
\eeq
If we take the trace of (\ref{ric1}) together with (\ref{ric2}),
we obtain
\beq
(\mbox{tr}{\cal L})^2 - \mbox{tr}({\cal L}^2) - R = 0~,
\eeq
which gives a constraint $E=T+V=0$ of this system.
Therefore, the trajectories of ``point particle'' living on the level set
$E=0$ represent Ricci-flat Riemannian manifolds. Introducing
a new time parameter $\tau$ defined by $dt=a^2 b^2 c^2 f d\tau$,
we can write the kinetic term as
\beq
T=\frac{1}{2}g_{ij}\frac{d\alpha^i}{d\tau} \frac{d\alpha^j}{d\tau},
\eeq
where the metric is given by
\beq
g_{ij}=\pmatrix{
4 & 8 & 8 & 4 \cr
8 & 4 & 8 & 4 \cr
8 & 8 & 4 & 4 \cr
4 & 4 & 4 & 0 \cr
},
\eeq
and $\alpha^i=(\alpha, \beta, \gamma, \sigma)$ with $a=e^{\alpha},
b=e^{\beta}, c=e^{\gamma}, f=e^{\sigma}$.
The potential $V$ is expressed in terms of a superpotential
$W$ as
\beq
V=-\frac{1}{2}g^{ij}\frac{\del W}{\del \alpha^i}
\frac{\del W}{\del \alpha^j}
\eeq
with
\beq
W=4abcf(a^2+b^2+c^2)+2f^2(n_1 b^2 c^2
+ n_2 a^2c^2+ n_3 a^2 b^2)~.
\eeq
Thus the Ricci-flatness condition follows from
the gradient flow equation,
\beq
\frac{d\alpha^i}{d\tau}=g^{ij}\frac{\del W}{\del \alpha^j}~,
\eeq
which reproduces the instanton equation (\ref{inst}).

\section{ ALC solutions}
\setcounter{equation}{0}

Let us first make a change of variable defined by $dr = f(t)dt$
and take the following ansatz to solve the instanton equation (\ref{inst});
\beqa
a^2 (r) &=& \frac{2 n_1}{\alpha_1 -
\alpha_2} (r - \alpha_1)(r -\alpha_2)~, \CR
b^2 (r) &=& \frac{2 n_2}{\beta_1 - \beta_2}
(r - \beta_1)(r -\beta_2)~, \label{ansatz} \\
c^2 (r) &=& \frac{2 n_3}{\gamma_1 -
\gamma_2} (r - \gamma_1)(r -\gamma_2)~, \nonumber
\eeqa
so that we have $a^2(r), b^2(r), c^2(r) \sim r^2$ as $r \to \infty$.
The overall normalizations are fixed by the requirement that we can make a
quadrature of the
differential equation for $f(r)$ to obtain
\beq
f^2(r) = \frac{(r - \alpha_1)(r - \beta_1)(r - \gamma_1)}
{(r - \alpha_2)(r - \beta_2)(r - \gamma_2)}~.
\eeq
Since $f(r) \sim {\rm constant}$ as $r \to \infty$ in our ansatz,
asymptotically there is an $S^1$ of a constant
radius at infinity. Thus solutions obtained by this ansatz will give ALC
(asymptotically locally conical) metrics
in the sense of \cite{CGLP3}. We find that if the parameters obey
\beq
\alpha_1 - \alpha_2 = 2 n_1~, \qquad
\beta_1 - \beta_2 = 2 n_2~, \qquad
\gamma_1 - \gamma_2 = 2 n_3~, \label{cond1}
\eeq
and
\beq
\alpha_1 + \beta_1 = 2\gamma_2~, \qquad
\beta_1 + \gamma_1 = 2\alpha_2~, \qquad
\gamma_1 + \alpha_1 = 2\beta_2~, \label{cond2}
\eeq
then the ansatz (\ref{ansatz}) gives a $Spin(7)$ gravitational instanton.
We have expressed the conditions in a completely $\Sigma_3$ symmetric manner.
Note that due to the constraint $n_1 + n_2 + n_3=0$, one
of the
six conditions is redundant and we have one free parameter that corresponds to
a translation of the radial coordinate $r$. After rescaling the radial
coordinate $r \rightarrow r/\ell$ by an arbitrary positive parameter $\ell$
with
dimensions of length, our ALC
solutions can be written as
\beqa
a^2(r) &=& (r-\alpha_1 \ell)(r-\alpha_2 \ell), \CR
b^2(r) &=& (r-\beta_1 \ell)(r-\beta_2 \ell), \CR
c^2(r) &=& (r-\gamma_1 \ell)(r-\gamma_2 \ell), \label{sol} \\
f^2(r) &=& \ell^2 \frac{(r-\alpha_1 \ell)(r-\beta_1 \ell)(r-\gamma_1 \ell)}
{(r-\alpha_2 \ell)(r-\beta_2 \ell)(r-\gamma_2 \ell)} \nonumber
\eeqa
with the conditions (\ref{cond1}) and (\ref{cond2}).
The asymptotic form of the metric is
\beq
g \sim dr^2 + r^2 d\Omega_6 + \ell^2 T_A^2~,
\eeq
where $d\Omega_6$ is a homogeneous metric on the flag manifold $SU(3)/U(1)
\times U(1)$,
which is the twister space of ${\bf{CP}}(2)$.

Let us look at a few special examples, where a cancellation of a zero and a
pole
of the rational function $f^2$ takes place.

\begin{enumerate}

\item \underline{$\overrightarrow n =
(1,-1,0)$}

In this case we can take $\gamma_1=\gamma_2=0$
by a translation of $r$.
Then the solution is
\beqa
a^2 (r) &=& (r - 4\ell/3)(r + 2\ell/3)~, \CR
b^2 (r) &=& (r + 4\ell/3)(r -2\ell/3), \label{exs} \\
c^2 (r) &=& r^2~, \CR
f^2(r) &=& \ell^2\frac{(r - 4\ell/3)(r + 4\ell/3)}
{(r + 2\ell/3)(r - 2\ell/3)}~. \nonumber
\eeqa
The solution has the same structure\footnote{
It is only at the level of solutions to the first order system
and does not mean the geometry is the same,
since the starting coset space is different.}
as new $G_2$ metrics in \cite{BGGG}
based on $S^3 \times S^3$. The metric is regular in the region $r > 4\ell/3$.
At the boundary $a^2 \to 0, f^2 \to 0$ but $b^2 = c^2 = 16\ell^2/9$.
Since $b^2$ and $c^2$ approach the same boundary value,
we have {\bf CP}(2) as a singular orbit (see the next section).

\item \underline{$\overrightarrow n =
(1,1,-2)$}

In this case $\alpha_i = \beta_i$ has to be satisfied and we obtain
a reduced solution with $a^2=b^2$. We take $\alpha_1=\beta_1= 1$.
Then the solution is
\beqa
a^2 (r) &=& (r - \ell)(r + \ell)~, \CR
b^2 (r) &=& a^2(r), \\
c^2 (r) &=& (r + 3\ell)(r - \ell)~, \CR
f^2(r) &=& \ell^2\frac{(r - \ell)(r + 3\ell)}
{(r + \ell)^2}~. \nonumber
\eeqa
We see the solution has the same form as the simplest solution
(denoted ${\bf A}_8$) among new $Spin(7)$ metrics in \cite{CGLP3}.
The metric is regular in the region $r > \ell$ and at $r=\ell$
all the coefficients are linearly vanishing. Thus the principal orbit
$SU(3)/U(1)$ collapses to a point and the manifold has curvature
singularities at $r=\ell$, since $SU(3)/U(1)$ is not homeomorphic to
$S^7$.

\item \underline{$\overrightarrow n =
(2, -1, -1)$}

This is obtained from the second example simply by the sign flip and
a permutation, but the global topology is different as we will see shortly.
In the same way as above the solution is
\beqa
a^2 (r) &=& (r - 3\ell)(r + \ell)~, \CR
b^2 (r) &=& (r - \ell)(r + \ell), \label{ex3} \\
c^2 (r) &=& b^2(r)~, \CR
f^2(r) &=& \ell^2\frac{(r + \ell)(r - 3\ell)}
{(r - \ell)^2}~. \nonumber
\eeqa
The regular region is $r > 3\ell$ and in contrast to the second example
we have  ${\bf CP}(2)$ with finite volume at the boundary.
This solution corresponds to the solution denoted ${\bf B}_8$ in \cite{CGLP3}.

\end{enumerate}

\section{Perturbation around the singular orbit}
\setcounter{equation}{0}

In this section we will give a perturbative expansion in a small neighborhood
of the singular orbit $(t=0)$ for the $Spin(7)$ gravitational instanton
equation (\ref{inst}).
A mathematical foundation may be found in \cite{EW}. Our metric is written
in the form $g=dt^2+g_t$, where $g_t \; (t \ge 0)$ is a one-parameter
family of
$SU(3)$-invariant metrics on the principal orbit $SU(3)/U(1)$.
We assume that near the boundary $t=0$ the orbit is locally of the form
\beq
SU(3)/U(1) \longrightarrow S^3 \times SU(3)/U(2)~,
\eeq
where $S^3$ denotes a round 3-sphere whose radius tends to zero at $t=0$,
and the singular orbit $SU(3)/U(2)$ is a complex projective
space ${\bf CP}(2)$ whose size remains non-vanishing at $t=0$. Thus, if
we choose the rate of collapse of the $S^3$ factor appropriately,
the manifold approaches ${\bf R}^4 \times {\bf CP}(2)$ at short distance; it
has topologically the same local behavior as the hyperk\"ahler manifold
$T^{*}{\bf CP}(2)$. In general
a singular orbit gives a singularity of the instanton equation
and we have no smooth solution at the
singularity. However, in our case, the very geometric nature of the equation
allows a smooth solution in a neighborhood around a singular orbit.

Due to the $\Sigma_3$-symmetry of our instanton equation,
we have three types of possible boundary
conditions for the limit $t \rightarrow 0$:
\beqa
g & \longrightarrow & dt^2+t^2(T_A^2/n_1^2+\sigma_1^2+\sigma_2^2)
+m^2(\Sigma_1^2+\Sigma_2^2+\tau_1^2+\tau_2^2), \label{bc1} \\
g & \longrightarrow & dt^2+t^2(T_A^2/n_2^2+\Sigma_1^2+\Sigma_2^2)
+m^2(\sigma_1^2+\sigma_2^2+\tau_1^2+\tau_2^2), \label{bc2}\\
g & \longrightarrow & dt^2+t^2(T_A^2/n_3^2+\tau_1^2+\tau_2^2)
+m^2(\sigma_1^2+\sigma_2^2+\Sigma_1^2+\Sigma_2^2), \label{bc3}
\eeqa
where $m$ is a scale parameter corresponding to the size of ${\bf CP}(2)$.
By choosing a boundary condition the $\Sigma_3$-symmetry is broken to
the ${\bf Z}_2$ symmetry.
For each choice of the boundary condition which specifies a ${\bf CP}(2)$
embedded in $SU(3)/U(1)$,
the unit volume element $\overrightarrow v_{\alpha} (\alpha=1,2,3)$ of
the ${\bf CP}(2)$ is given by
\beq
\overrightarrow v_{1}=e^3 \wedge e^4 \wedge e^5 \wedge e^6, \quad
\overrightarrow v_{2}=e^1 \wedge e^2 \wedge e^5 \wedge e^6, \quad
\overrightarrow v_{3}=e^1 \wedge e^2 \wedge e^3 \wedge e^4,
\eeq
respectively. For all three cases
the calibration $\Omega$ given by (\ref{calib}) satisfies the equation
$|\Omega(\overrightarrow v_{\alpha})|=1$.
We therefore see that the singular orbit ${\bf CP}(2)$ appearing at the
boundary is a Cayley submanifold (supersymmetric four-cycles) in
$Spin(7)$ holonomy manifold.

The solutions with these three boundary conditions and the corresponding
${\bf CP}(2)$'s are permuted by the action of $\Sigma_3$.
For concreteness, we will consider from now
on the first boundary condition (\ref{bc1}).
The quantity $\Sigma_1^2+\Sigma_2^2+\tau_1^2+\tau_2^2$ represents
the Fubini-Study metric
on ${\bf CP}(2)$ and $T_A^2/n_1^2+\sigma_1^2+\sigma_2^2$ locally
the metric on the unit 3-sphere. More precisely the metric $g$ would
have an orbifold
singularity at $t=0$ unless we choose the value of $\overrightarrow n$
appropriately,
since the latter represents the metric on the Lens space $S^3/{\bf Z}_n$
in general rather than $S^3$ globally (see the next section).
The perturbative series expansion around the singular orbit ${\bf CP}(2)$
yields
\beqa
a(t) & = & t \left(1-\frac{1}{2}(Q+1)(t/m)^2+ \cdot \cdot \cdot \right), \CR
b(t) & = & m \left(1+\frac{1}{6}(4-n_2/n_1)(t/m)^2+ \cdot \cdot
\cdot
\right), \label{local} \\
c(t) & = & m \left(1+\frac{1}{6}(4-n_3/n_1)(t/m)^2+ \cdot \cdot
\cdot
\right), \CR
f(t) & = & \frac{t}{n_1}\left(1+Q(t/m)^2+
\cdot \cdot \cdot \right). \nonumber
\eeqa
It should be noticed that the solution is not uniquely determined by the
boundary condition and it includes an additional free parameter
$Q$. The expansion (\ref{local}) is a consequence of the assumption (\ref{bc1})
and valid for any solution that is smooth around the singular orbit.
However, it is not at all clear that the local perturbative solution
(\ref{local})
can be extended to a global complete metric. There should be some bound
on $Q$ that may depend on a choice of $\overrightarrow n$.
One can obtain an example of local solution that extends
to a complete metric by setting
$\overrightarrow n = (1,1,-2)$\footnote{With this choice of $\overrightarrow n$
the ALC solution cannot have a singular orbit with finite volume
   (see the second example in section 3).}  and $Q=-2/3$.
Then it can be extended to the Calabi
hyperk\"ahler metric on $T^{*}{\bf CP}(2)$ of $Sp(2)$
holonomy \cite{CGLP2},\cite{DS};
\beq
a^2(r)=\frac{1}{2}(r^2-m^2), \quad b^2(r)=\frac{1}{2}(r^2+m^2), \quad
c^2(r)=r^2, \quad f^2(r)=\frac{r^2}{4}(1-(m/r)^4)
\eeq
with $dt=dr/(1-(m/r)^4)^{1/2}$.
Note that the asymptotic behavior of the Calabi metric at infinity is
different from the ALC metrics.

We can also find the specialization
which reproduces the ALC metrics described in the previous section.
Let us consider the solution (\ref{sol}) whose radial coordinate
is constrained to be $r\ge \alpha_1 \ell$. From the conditions
(\ref{cond1}) and (\ref{cond2}) we have
\beqa
\alpha_1 - \alpha_2 &=& 2 n_1, \quad \alpha_1 - \beta_1 =
\frac{4}{3}(n_1 - n_2), \quad \alpha_1 - \beta_2=
\frac{2}{3}(n_1 - n_3), \CR
\alpha_1 - \gamma_1 &=& \frac{4}{3}(n_1 - n_3),
\quad \alpha_1 - \gamma_2 = \frac{2}{3}(n_1 - n_2).
\eeqa
Thus if we choose the parameter $\overrightarrow n$ as
\beq
n_1 > 0, \quad n_1 > n_2, \quad n_1 >
n_3, \label{ineq}
\eeq
then the solution $\{a,b,c,f\}$ is non-vanishing in the region
$r > \alpha_1 \ell$, and so the ALC metric is non-singular.
The behavior of the metric near $r=\alpha_1 \ell$ is given by
\beq
g \longrightarrow d\rho^2 + \rho^2(T_A^2/n_1^2
+\sigma_1^2+\sigma_2^2)+\ell^2(\alpha_1 - \beta_1)
(\alpha_1 - \beta_2)(\Sigma_1^2+\Sigma_2^2+\tau_1^2+\tau_2^2),
\eeq
where $\rho^2=\ell(\alpha_1 - \alpha_2)(r-\alpha_1 \ell)$.
Setting $m=\ell\sqrt{(\alpha_1 - \beta_1)(\alpha_1 - \beta_2)}$,
we reproduce the equation (\ref{bc1}) and higher order
calculations yield the relation
\beq
Q = -\frac{2}{27}(13+2n_2 n_3/n_1^2).
\eeq

The parameter in the perturbative solution $Q$ implies a possibility
of non-trivial deformations of the ALC solutions (\ref{sol}).
Although we have not been able to find general solutions in closed form,
numerical simulations of the instanton equation indicate a family
of global solutions under some conditions of $Q$.
For example, in the case of $\overrightarrow n=(1,-1,0)$,
the condition is given by $Q \le -0.35$ approximately and the
exact solution (\ref{exs}) with $Q=-26/27$ is included in
this region.  The existence of more general solutions is also
supported by a similar analysis for the coset space $Sp(2)/Sp(1)$,
where the general solutions of complete metric have been
obtained \cite{CGLP3}. In Appendix B it is shown briefly
how we can accommodate a local perturbative analysis around
the singular orbit of $Sp(2)/Sp(1)$ model to the global solutions in
\cite{CGLP3}.

\section{The issue of global topology}
\setcounter{equation}{0}

Let us consider the global topology of solutions with boundary condition
(\ref{bc1})
by calculating explicitly the 1-forms $\sigma_1, \sigma_2$ and $T_A$
describing the metric on the unit 3-sphere $S^3$ locally.
Our calculation shows that near the boundary the topology of the principal
orbit is in general $S^3/{\bf Z}_{n_1} \times {\bf CP}(2)$ rather than $S^3
\times {\bf CP}(2)$. The integer $n_1$ comes in here since we take the
first boundary condition (\ref{bc1}). For other boundary conditions (\ref{bc2})
and (\ref{bc3}), $n_1$ is replaced by $n_2$ or $n_3$ accordingly.

To consider the topology of the fiber over a point of the base space ${\bf
CP}(2)$
we fix the coordinates on ${\bf CP}(2)$. Then $\Sigma_i=\tau_j=0$ on the
fiber and
the $SU(3)$ Maurer-Cartan equation reduces to the following form;
\beqa
d\sigma_1 &=& \kappa_A T_A \wedge \sigma_2 + \kappa_B T_B \wedge \sigma_2, \CR
d\sigma_2 &=& -\kappa_A T_A \wedge \sigma_1
- \kappa_B T_B \wedge \sigma_1, \label{redMC} \\
dT_A &=& 2\alpha_A \sigma_1 \wedge \sigma_2, \CR
dT_B &=& 2\alpha_B \sigma_1 \wedge \sigma_2.  \nonumber
\eeqa
In fact this Maurer-Cartan equation
comes from the following choice of
the generators $E_{\alpha} (\alpha=1,2$ or $A,B)$ of $U(2)$;
\beq
E_1 = \left(
\begin{array}{ccc}
0 & 0 & 0 \\
0 & 0 & 1 \\
0 & 1 & 0
\end{array}
\right),
E_2 = \left(
\begin{array}{ccc}
0 & 0 & 0 \\
0 & 0 & -i \\
0 & i & 0
\end{array}
\right),
E_A = -\frac{1}{\Delta} \left(
\begin{array}{ccc}
\alpha_B & 0 & 0 \\
0 & \beta_B & 0 \\
0 & 0 & \gamma_B
\end{array}
\right),
E_B=\frac{1}{\Delta} \left(
\begin{array}{ccc}
\alpha_A & 0 & 0 \\
0 & \beta_A & 0 \\
0 & 0 & \gamma_A
\end{array}
\right). \label{base}
\eeq
By expanding a left invariant one form $\omega$ of the subgroup $U(2)
\subset SU(3)$;
\beq
\omega = i(\sigma_1 E_1+\sigma_2 E_2+T_A E_A+T_B E_B), \label{mat}
\eeq
and using the relations in Appendix A,
we can check the equation $d\omega + \omega \wedge \omega = 0$
implies (\ref{redMC}).

The left invariant one form $\omega$ is represented by $\omega = g^{-1}dg$
in terms of
a group element $g \in U(2)$. To parametrize the group element $g$
we use Euler angles $(\theta,\phi,\psi)$ with the range
\beq
0\le\theta<\pi, \quad 0\le\phi<2\pi, \quad 0\le\psi<4\pi
\eeq
of $SU(2)$ and a U(1) coordinate $\chi$ defined by
\beq
U_B(1)=\exp(i\Delta \chi E_B/2)=\mbox{diag}
(e^{i n_1 \chi/2}, e^{i n_2 \chi/2},
e^{i n_3 \chi/2}),
\eeq
where $(n_1, n_2, n_3):=
(\alpha_A, \beta_A, \gamma_A)$ are integers
with no common divisor\footnote{See the discussion
below eq. (\ref{inst}). } satisfying
the traceless condition $n_1 +n_2 +n_3=0$.
With these coordinates the group element $g$ is given by
\beqa
g(\phi,\theta,\psi,\chi) &=&
e^{i(\phi/2)E_3}e^{i(\theta/2)E_2}e^{i(\psi/2)E_3}
e^{i(\Delta \chi/2)E_B} \CR
&=& \pmatrix{
e^{i n_1 \chi/2} & 0 & 0 \cr
0 & \cos(\theta/2)e^{i(\psi + \phi + n_2 \chi)/2}
& \sin(\theta/2)e^{-i(\psi - \phi - n_3 \chi)/2} \cr
0 & -\sin(\theta/2)e^{i(\psi - \phi + n_2 \chi)/2}
& \cos(\theta/2)e^{-i(\psi + \phi - n_3 \chi)/2} \cr
},
\eeqa
where
\beq
E_3 = \alpha_A E_A + \alpha_B E_B
           = \left(
\begin{array}{ccc}
0 & 0 & 0 \\
0 & 1 & 0 \\
0 & 0 & -1
\end{array}
\right).
\eeq
We now obtain the one-forms from (\ref{mat}),
\beqa
\sigma_1 &=& \frac{1}{2} \sin\theta \cos(\psi+(n_2-n_3)\chi/2)
d\phi - \frac{1}{2} \sin(\psi+(n_2-n_3)\chi/2)d\theta, \CR
\sigma_2 &=& \frac{1}{2} \sin\theta \sin(\psi+(n_2-n_3)\chi/2)
d\phi + \frac{1}{2} \cos(\psi+(n_2-n_3)\chi/2)d\theta, \\
T_A &=& \frac{n_1}{2}(d\psi + \cos\theta d\phi). \nonumber
\eeqa
These one-forms give a metric
\beq
T_A^2/n_1^2 + \sigma_1^2 + \sigma_2^2 = \frac{1}{4}(d\theta^2 +
\sin^2 \theta d\phi^2 ) + \frac{1}{4}(d\psi + \cos\theta d\phi)^2 \label{s3}
\eeq
on the coset space $U(2)/U_B(1)$ whose topology is locally
$S^3$, but not globally $S^3$. In fact for each integer $p=0,1,\cdots, n_1-1$
we can find an integer $q$ such that
we have
\beq
g(\phi,\theta,\psi+4\pi p/n_1,\chi+4\pi q/n_1) =
g(\phi,\theta,\psi,\chi).
\eeq
Thus the angle $\psi$ is identified with $\psi+4\pi p/ n_1$,
which implies the equation (\ref{s3}) expresses the metric
on $S^3/{\bf Z}_{n_1}$. This is the total space
of $U(1)$-bundle over $S^2$ with a connection
(Wu-Yang monopole potential) $T_A$ and the topological index
is given by \cite{QM}
\beq
\frac{1}{2\pi} \int_{S^2} dT_A = n_1.
\eeq

In conclusion, if we normalize $\overrightarrow n=(n_1,n_2,n_3)$
as integers with no common divisor,
then global solutions to the instanton equation (\ref{inst}) with the boundary
condition (\ref{bc1}) describe
manifolds of special holonomy which behave like
${\bf R}^4/{\bf Z}_{n_1} \times {\bf CP}(2)$ near the singular orbit.
For example the Calabi metric which has $\overrightarrow n=(1, 1, -2)$
is a hyperk\"ahler metric on $T^* {\bf CP}(2)$ with a fiber ${\bf C}^2 =
{\bf R}^4$.
Concerning the examples of ALC solutions in section 3,
the solution (\ref{exs}) with $\overrightarrow n=(1, -1, 0)$
has a trivial fiber ${\bf R}^4$, while the fiber of the solution (\ref{ex3})
with $\overrightarrow n=(2, -1, -1)$ is ${\bf R}^4/{\bf Z}_2$.
By the constraint (\ref{ineq}) on $\overrightarrow n$,
among our ALC solutions (\ref{sol}) only the solution (\ref{exs})
gives a complete metric of $Spin(7)$ holonomy
without orbifold singularities.

\section{$L^2$-normalisable harmonic 4-forms}
\setcounter{equation}{0}

In this section we consider $L^2$-normalisable harmonic
4-forms on ALC manifolds of $Spin(7)$ holonomy.
For the metric (\ref{metric}) of cohomogeneity one, we assume
the following self-dual 4-form
$G$~;
\beqa
G &=& u_{1}(t)(e^{0567}+e^{1243})+u_{2}(t)(e^{0473}+e^{2561})
+u_{3}(t)(e^{0127}+e^{3654}) \CR
&+& u_{4}(t)(e^{0315}+e^{4267}+e^{0524}+e^{6371}+e^{0461}+e^{7325}
+e^{0362}+e^{5471}),
\eeqa
where $e^{abcd}=e^a \wedge e^b \wedge e^c \wedge e^d$.
If $u_{A}=1$ for all $A$, then the 4-form $G$ is equal to the
calibration $\Omega$.
The closedness condition $dG=0$ is expressed by the
equations,
\beqa
\frac{d}{dt}(a^2 b^2 u_1) &=& 4abc u_{4} - 2n_1 fb^2 u_{2}
-2n_2 fa^2 u_{3}, \CR
\frac{d}{dt}(a^2 c^2 u_2) &=& 4abc u_{4} - 2n_1 fc^2 u_{1}
-2n_3 fa^2 u_{3}, \label {closed} \\
\frac{d}{dt}(b^2 c^2 u_3) &=& 4abc u_{4} - 2n_2 fc^2 u_{1}
-2n_3 fb^2 u_{2}, \CR
\frac{d}{dt}(abcf u_4) &=& f(a^2 u_3 + b^2 u_2 + c^2 u_1). \nonumber
\eeqa
The instanton equation leads to a linear relation as the first integral
\beq
u_1 + u_2 + u_3 + 4 u_4 = k, \label {rel}
\eeq
where the constant $k$ should be chosen to zero by the
$L^2$-normalisability of the 4-form $G$.
This choice is also consistent with the criterions for
unbroken supersymmetry\footnote{We thank Katrin Becker
for pointing out an error in the original version.} for 
compactifications of $M$ theory
on manifolds of $Spin(7)$ holonomy \cite{BK};
\beq
*G=G, \quad \omega=0, \quad G \wedge \Omega =0,
\eeq
where $\omega$ is the 2-form defined by 
$\omega = G_{abcd} \Psi_{abcf} e^d \wedge e^f$.
In fact our ansatz satisfies $*G=G,~\omega=0 $ automatically 
and the last equation 
requires precisely the relation
(\ref{rel}) with $k=0$.
Thus following \cite{CGLP1},\cite{CGLP2},\cite{CGLP3},
one can construct supersymmetric
M2-branes using the $L^2$-normalisable solutions of (\ref{closed}).

In order to obtain explicit solutions of (\ref{closed}),
we take the ALC solution (\ref{exs}) as the background metric.
Then the radial coordinate $r$ runs from the singular orbit at
$r=(4/3)\ell$ to infinity.
It is convenient to introduce
the new variables~,
\beq
u=b^2c^2 u_{3}-a^2c^2 u_{2}, \quad  v=b^2c^2 u_{3}+a^2c^2 u_{2},
\quad w=a^2b^2 u_{1}.
\eeq
After eliminating $u_4$ by (\ref{rel}) and further taking
derivatives of the first-order equations (\ref{closed}),
we obtain the
Fuchs type differential equation\footnote{We set the length parameter
$\ell$ in the solution (\ref{exs}) to unity for convenience.};
\beq
\frac{d^3}{dr^3}u + p_{1}(r) \frac{d^2}{dr^2}u + p_{2}(r)
\frac{d}{dr}u + p_{3}(r) u = 0,
\eeq
where
\beqa
p_{1}(r) &=& \frac{8(2560 + 12384 r^2 - 29160 r^4 + 10206 r^6 + 6561 r^8)}
{81r(r + 2/3)(r - 2/3)(r + 4/3)(r - 4/3)(9r^2 + 20)(9r^2 - 8)}, \CR
p_{2}(r) &=& \frac{2(-10240 + 26496 r^2 - 116640 r^4 + 72900 r^6 + 32805 r^8)}
{81r^2(r + 2/3)(r - 2/3)(r + 4/3)(r - 4/3)(9r^2 + 20)(9r^2 -8)}, \CR
p_{3}(r) &=& -\frac{48r(9r^2 + 104)}{(r + 2/3)(r - 2/3)(r + 4/3)(r - 4/3)
(9r^2 + 20)(9r^2 - 8)}.
\eeqa
This equation can be integrated by imposing the regularity of the solution
at the regular singular point $r=4/3$
and we find a solution
\beq
u(r)=\frac{27r^2(9r^2 - 40)}{(r + 2/3)(r - 2/3)(r + 4/3)^2}. \label{fuchs}
\eeq
The remaining functions $v$ and $w$ are given by taking derivatives
of (\ref{fuchs}). Finally we
obtain an $L^2$-normalisable harmonic 4-form
$G$ in the region $r \ge 4/3$~;
\beqa
u_{1} &=& \frac{2(160-72r^2+243r^3+81r^4)}
{r(r+2/3)^2(r-2/3)^2(r+4/3)^3}, \CR
u_{2} &=& -\frac{3(32+48r-72r^2+135r^3)}
{2r^3(r-2/3)(r+2/3)^2(r+4/3)}, \\
u_{3} &=& \frac{512+768r-1440r^2-4752r^3+648r^4+243r^5}
{6r^3(r+2/3)(r-2/3)^2(r+4/3)^3}, \CR
u_{4} &=& \frac{64+144r+180r^2-81r^3}{r^3(r+2/3)(r-2/3)(r+4/3)^3}. \nonumber
\eeqa


\vskip10mm

\begin{center}
{\bf Acknowledgements}
\end{center}

We would like to thank Y. Hashimoto, M. Naka, H. Ohta and T. Ootsuka
for useful discussions. We would also like to thank T. Eguchi for
discussions and a helpful comment.
This work is supported in part by the fund
for special priority area 707
"Supersymmetry and Unified Theory of Elementary Particles" and
the Grant-in-Aid for Scientific Research No. 12640074.


\begin{flushleft}
{\bf Note Added}
\end{flushleft}

After we submitted this paper to e-prints archives,
we noticed a new paper \cite{CGLP5}, which has
a considerable overlap with our paper.
Many of our results have been also obtained
in section 3 of \cite{CGLP5} and we find a complete agreement.
Later we also received another preprint \cite{GS} which studies
$M$ theory on $Spin(7)$ manifolds. In \cite{GS} the regular solution
obtained as a special case of our ALC solutions is constructed 
independently and
employed in $M$ theory compactification.

\section*{Appendix A}
\renewcommand{\theequation}{A.\arabic{equation}}\setcounter{equation}{0}

The spin connection on the coset space $SU(3)/U(1)$ is obtained
by the Maurer-Cartan equation for the left-invariant one forms of $SU(3)$.
To derive the instanton equation in $\Sigma_3$ symmetric form, we here present
the $\Sigma_3$ symmetric $SU(3)$ Maurer-Cartan equation;
\beqa
d\sigma_1 &=& \Sigma_1 \wedge \tau_1 - \Sigma_2 \wedge \tau_2 + \kappa_A
T_A \wedge \sigma_2
+ \kappa_B T_B \wedge \sigma_2~, \CR
d\sigma_2 &=& - \Sigma_1 \wedge \tau_2 - \Sigma_2 \wedge \tau_1 - \kappa_A
T_A \wedge \sigma_1
- \kappa_B T_B \wedge \sigma_1~, \CR
d\Sigma_1 &=& \tau_1 \wedge \sigma_1 - \tau_2 \wedge \sigma_2 + \mu_A T_A
\wedge \Sigma_2
+ \mu_B T_B \wedge \Sigma_2~, \CR
d\Sigma_2 &=& - \tau_1 \wedge \sigma_2 - \tau_2 \wedge \sigma_1 - \mu_A T_A
\wedge \Sigma_1
- \mu_B T_B \wedge \Sigma_1~, \\
d\tau_1 &=& \sigma_1 \wedge \Sigma_1 - \sigma_2 \wedge \Sigma_2 + \nu_A T_A
\wedge \tau_2
+ \nu_B T_B \wedge \tau_2~, \CR
d\tau_2 &=& - \sigma_1 \wedge \Sigma_2 - \sigma_2 \wedge \Sigma_1 - \nu_A
T_A \wedge \tau_1
- \nu_B T_B \wedge \tau_1~, \CR
dT_A &=& 2\alpha_A \sigma_1 \wedge \sigma_2  + 2\beta_A \Sigma_1 \wedge
\Sigma_2
+ 2\gamma_A \tau_1 \wedge \tau_2~, \CR
dT_B &=& 2\alpha_B \sigma_1 \wedge \sigma_2  + 2\beta_B \Sigma_1 \wedge
\Sigma_2
+ 2\gamma_B \tau_1 \wedge \tau_2~. \nonumber
\eeqa
This form of the Maurer-Cartan equation is symmetric under the (cyclic)
permutation of $(\sigma_i, \Sigma_i,
\tau_i)$\footnote{To compare with the one in \cite{CGLP2} , we make an exchange
$\Sigma_1 \leftrightarrow \Sigma_2$
and a change of notation and sign $(\nu_1, \nu_2) \to (\tau_1, -\tau_2)$.}.
We have introduced parameters
$\alpha, \beta, \gamma, \kappa, \mu,\nu$, which describe the "coupling" of
the Cartan generators $\{ T_A, T_B\}$.
The condition $d^2 = 0$, or the Jacobi identity implies the following
constraints for them;
\beqa
\alpha_A + \beta_A + \gamma_A =
\alpha_B + \beta_B + \gamma_B &=& 0~, \CR
\kappa_A + \mu_A + \nu_A =
\kappa_B + \mu_B + \nu_B &=& 0~, \CR
\kappa_A \beta_A + \kappa_B \beta_B =\kappa_A \gamma_A + \kappa_B \gamma_B
&=& 1~, \\
\mu_A \gamma_A + \mu_B \gamma_B =\mu_A \alpha_A + \mu_B \alpha_B &=& 1~, \CR
\nu_A \alpha_A + \nu_B \alpha_B =\nu_A \beta_A + \nu_B \beta_B &=& 1~.
\nonumber
\eeqa
These equations can be solved in the form,
\beqa
\kappa_A &=& \frac{1}{\Delta}(\beta_B - \gamma_B), \;
\kappa_B = -\frac{1}{\Delta}(\beta_A - \gamma_A), \;
\mu_A = -\frac{1}{\Delta}(\alpha_B-\gamma_B), \CR
\mu_B &=& \frac{1}{\Delta}(\alpha_A - \gamma_A), \;
\nu_A = \frac{1}{\Delta}(\alpha_B - \beta_B), \;
\nu_B = -\frac{1}{\Delta}(\alpha_A - \beta_A)
\eeqa
with $\Delta=\beta_A \alpha_B -\alpha_A \beta_B$
leaving four free parameters $(\alpha_{A,B}, \beta_{A,B})$.
We may further put the "orthogonality" conditions;
\beqa
\alpha_A \alpha_B + \beta_A \beta_B + \gamma_A \gamma_B &=& 0~, \CR
\kappa_A \kappa_B + \mu_A \mu_B + \nu_A \nu_B &=& 0~,
\eeqa
which reduces one parameter.
A standard choice of parameters (cf.\cite{CGLP2}) is
\beqa
(\alpha_A, \beta_A, \gamma_A) &=& (1,1,-2)~,
\quad (\alpha_B, \beta_B, \gamma_B) = (1, -1, 0)~, \CR
(\kappa_A, \mu_A, \nu_A) &=& (-1/2,-1/2,1)~,
\quad (\kappa_B, \mu_B, \nu_B) =  (-3/2, 3/2, 0)~,
\eeqa
and the remaining three parameters correspond to the two scalings of
each $T_{A,B}$ and the overall rotation.

\section*{Appendix B}
\renewcommand{\theequation}{B.\arabic{equation}}\setcounter{equation}{0}

In this appendix we present a perturbative analysis of the $Spin(7)$
gravitational instanton equation based on the coset space $Sp(2)/Sp(1)$
and discuss the relation to the exact regular solutions constructed in
\cite{BS},\cite{GPP},\cite{CGLP3}.
Let us write the $Spin(7)$ metric in the form
\beq
g = dt^2 + a(t)^2(\Sigma_{1}^{2}+\Sigma_{2}^{2})
+b(t)^2 \sigma^2 + c(t)^2(\tau_{1}^{2}+\tau_{2}^2+\tau_{3}^{2}+\tau_{4}^2).
\eeq
Here $\{\Sigma_{a},\sigma,\tau_{i}\}$ are one-forms defined by using the
Maurer-
Cartan forms of $Sp(2)$ \cite{CGLP3} and $\{a,b,c\}$ are functions of the
radial
variable $t$ associated with the decomposition of the isotropy
representation $sp(2)/sp(1)={\bf p}_1 \oplus {\bf p}_2 \oplus {\bf p}_3$
with dimensions 2, 1 and 4, respectively. The instanton equation
is given by
\beq
{\dot a}=1-\frac{b}{2a}-\frac{a^2}{c^2}, \quad {\dot b}=\frac{b^2}{2a^2}
-\frac{b^2}{c^2}, \quad {\dot c}=\frac{a}{c}+\frac{b}{2c}. \label{inst2}
\eeq

We now assume that near $t=0$ the principal orbit is locally of the form
\beq
Sp(2)/Sp(1) \longrightarrow S^3 \times S^4,
\eeq
where $S^3$ collapses as $t \to 0$, while $S^4$ remains with a finite
radius $m$
at $t=0$. Then on the singular orbit the boundary condition for the metric
functions can be written as
\beq
a(t) \rightarrow t/2, \quad b(t) \rightarrow t/2, \quad c(t) \rightarrow m
\label{bc4}
\eeq
for $t \rightarrow 0$. The perturbative solution of (\ref{inst2}) with the
boundary
condition (\ref{bc4}) is
\beqa
a(t) &=& \frac{t}{2} \left(1 + \frac{1}{2} \tilde{Q} (t/m)^2 + \cdot
\cdot \cdot  \right), \CR
b(t) &=& \frac{t}{2} \left(1 - \frac{1}{2} (2 \tilde{Q}+1)(t/m)^2+ \cdot
\cdot \cdot \right), \\
c(t) &=& m \left(1+\frac{3}{8}(t/m)^2+ \cdot \cdot \cdot \right). \nonumber
\eeqa
This solution includes a free parameter $\tilde{Q}$ in addition
to the scale parameter $m$. Thus the structure is very similar to
that of the solution (\ref{local}), although the geometrical setting is
different.
Fortunately, in this case, all regular $Spin(7)$ metrics are known
in the closed
form \cite{BS},\cite{GPP},\cite{CGLP3} and so we can read the condition
for $\tilde{Q}$ admitting
global solutions.
By making the power series expansion of the known solutions at $t=0$
we find that the global solutions can arise in the parameter region
$\tilde{Q} \ge -1/3$: the perturbative solutions lift to
the metrics of $Spin(7)$ holonomy defined on the bundle
of chiral spinors over $S^4$. More precisely the solutions
in the regions (a) $-1/3 < \tilde{Q} < 0$, (b) $\tilde{Q}=0$,
(c) $\tilde{Q}>0$ lift to the ALC metrics {\bf B}$_{8}^{+}$,
{\bf B}$_{8}$ and {\bf B}$_{8}^{-}$ respectively, using
the notation of \cite{CGLP3}.
The boundary $\tilde{Q}=-1/3$ corresponds to the metric
of $Spin(7)$ holonomy  obtained in \cite{BS},\cite{GPP},
and the limit $\tilde{Q} \rightarrow \infty$ reduces to the
metric of $G_2$ holonomy on the ${\bf R}^3$
bundle over $S^4$ \cite{BS},\cite{GPP}.
The ALC solutions in this paper corresponds to the solution {\bf B}$_{8}$
and we expect there are deformations like {\bf B}$_{8}^{+}$,
and {\bf B}$_{8}^{-}$ in our case too.


\end{document}